\documentclass{PoS}
\usepackage{amsmath}
\usepackage{amssymb}

\title{Quark number susceptibility at finite density and low temperature}

\ShortTitle{Quark number susceptibility at finite density and low temperature}

\author{\speaker{Pietro Giudice}
         \thanks{
           This work is carried as part of the UKQCD collaboration and
           the DiRAC Facility jointly funded by STFC, the Large Facilities 
           Capital Fund of BIS and Swansea University.
           We thank the DEISA Consortium (www.deisa.eu), 
           funded through the EU FP7
           project RI-222919, for support within the DEISA Extreme Computing
           Initiative. 
           JIS is funded by Science Foundation Ireland grant 11/RFP/PHY3193.}\\
  Department of Physics, College of Science, Swansea University, \\
  Singleton Park, Swansea SA2 8PP, United Kingdom\\
  E-mail: \email{p.giudice@swansea.ac.uk}}

\author{Simon Hands\\
  Department of Physics, College of Science, Swansea University, \\
  Singleton Park, Swansea SA2 8PP, United Kingdom\\
  E-mail: \email{S.J.Hands@swansea.ac.uk}}

\author{Jon-Ivar Skullerud\\
  Department of Mathematical Physics, National University of Ireland, \\
  Maynooth, County Kildare, Ireland \\
  E-mail: \email{jonivar@skullerud.name}}

\abstract
{
We study the quark number susceptibility in SU(2) lattice gauge theory with
two Wilson quark flavours at non-zero chemical potential and low
temperature. We present some technical aspects of the issue and numerical
results obtained at different lattices and different parameters. We discuss
what this observable can teach us about the phase diagram of the model and
in particular about the relation between susceptibility and Polyakov loop.
}

\FullConference{XXIX International Symposium on Lattice Field Theory \\
                 July 10 - 16, 2011\\
                 Squaw Valley, Lake Tahoe, California}

\newcommand{\beq}{\begin{equation}}
\newcommand{\eeq}{\end{equation}}
\newcommand{\bea}{\begin{eqnarray}}
\newcommand{\eea}{\end{eqnarray}}

\begin{document}

\section{Introduction}

Studying the statistical fluctuations of a system is a powerful 
method to characterise the thermodynamic properties of a system.
As a matter of fact, the presence of a phase transition is signalled 
by an enhancement of the fluctuations in the system.
The observables appropriate to extract this kind of information are called
susceptibilities. Properly, the susceptibility describes the response of 
a system to an applied field. Given an operator we can usually determine
a corresponding susceptibility that is the second derivative of the free 
energy density.
In this paper we are interested in studying the quark number susceptibility 
(QNS) $\chi=\frac{\partial n_q}{\partial \mu}$: it is the response of $n_q$, 
the quark number density, to an infinitesimal change in the quark chemical 
potential $\mu$.

The QNS has been studied so far at finite temperature and zero chemical 
potential, see Refs.~\cite{Gottlieb:1987ac, Allton:2005gk, Gavai:2008zr}.
This is largely because QNS is directly related to experimental 
measurements of fluctuations observed in heavy ion 
collisions~\cite{Asakawa:2000wh}. 
At non-zero density and zero (or low) temperature regime, QNS is not commonly
studied; an application to QCD of rainbow approximation of the 
Dyson-Schwinger approach to study the QNS can be found in 
Ref.~\cite{He:2008zzb}.  One reason for the lack of studies in this
regime is the well known sign problem: it is not possible, using
standard tools, to simulate lattice QCD at non-zero density.

In this paper we study the QNS at non-zero density and at low temperature 
in the context of $SU(2)$ gauge theory, {\it i.e.} where simulations 
are feasible.

\section{Two color QCD}

In two color QCD, {\it i.e.} $SU(2)$ gauge theory, quarks and antiquarks
live in equivalent representations of the group; the physical consequence is 
that $q\bar{q}$ mesons and $qq$, $\bar{q}\bar{q}$ baryons are contained in the 
same hadron multiplet. In the limit when $m_\pi \ll m_\rho$, {\it i.e.}
when the pion mass is very small compared with the first non-Goldstone hadron,
it is possible to study the system using the chiral perturbation theory ($\chi$PT)
limit~\cite{Kogut:2000ek}. The fundamental result is that only for
$\mu \geq \mu_0 \equiv \frac{1}{2} m_\pi$ the quark number density $n_q$ becomes
different from zero and at the same onset value $\mu_0$ also a condensate
$\langle qq \rangle  \neq 0$  develops, signalling the spontaneous breakdown 
of the global $U(1)$ baryon number symmetry: a superfluid phase appears.
In this phase there are tightly bound scalar diquarks but the hadrons are
weakly interacting between them:
therefore, just above the onset, the system is very dilute and is
described as a Bose Einstein Condensate (BEC).

Using $\chi$PT it is possible to determine the behaviour of different 
observables; in particular, we can write down the prediction for $n_q$ and
its susceptibility $\chi$, at zero temperature $T=0$ and in the limit 
$\mu  \to  \mu_0$ (right-hand limit):
\beq
n^{{\chi}PT}_q \approx 32 N_f F^2  \left( \mu - \mu_0 \right)
\ , \quad
\chi^{{\chi}PT} \approx 32 N_f F^2 -192 N_f \frac{F^2}{m_\pi} \left( \mu - \mu_0 
\right)
\ ,
\label{nchinCPT}
\eeq
where $F$ is the pion decay constant and $N_f$ the number of fermions. 
The diquark condensate, calculated in the same limits,  
is given in terms of the chiral condensate at zero chemical potential:
\beq
\langle q q \rangle \approx 
2 \langle \bar{q}q \rangle_0 \sqrt{\frac{\mu}{\mu_0} - 1} \ .
\label{diqCPT}
\eeq

There are reasons to think that above a certain 
value of the chemical potential, $\mu \ge \mu_Q$, a degenerate system of 
weakly interacting quarks is more stable~\cite{Hands:2006ve}. 
For an ideal gas (Stefan Boltzmann (SB) limit) of massless quarks and gluons, 
at $T=0$, we have:
\beq
n^{SB}_q=\frac{N_f N_c}{3 \pi^2} \mu^3 \ ,
\quad
\chi^{SB}=\frac{N_f N_c}{\pi^2} \mu^2 \ ,
\label{nchinSB}
\eeq
where $N_c$ is the number of colours.
In this case the superfluidity is explained by a BCS condensation of Cooper 
pairs within a layer of thickness $\Delta$ around the Fermi surface; the
diquark condensate is therefore given by: 
\beq
\langle qq \rangle  \propto \mu^2 \Delta \ .
\label{diqSB}
\eeq
Comparing Eq.s~(\ref{nchinCPT}),~(\ref{diqCPT}) with 
Eq.s~(\ref{nchinSB}),~(\ref{diqSB}), we see clearly that the two phases 
are characterised by two quite different behaviours.

We consider also the order parameter related to the confinement 
property of the theory: the Polyakov loop $L$; as discussed in 
Ref.~\cite{Hands:2010gd} the theory becomes deconfined only after 
$\mu \ge \mu_D$. Surprisingly, there is a regime, $\mu_Q < \mu < \mu_D$,
where the theory is confined, {\it i.e.} $\langle L \rangle = 0$, 
but the other observables seem to suggest non interacting fermions: 
a confined BCS phase. This phase could be the so called quarkyonic phase 
introduced in Ref.~\cite{McLerran:2007qj}; arguments against the extension
of this idea to the case of $N_c=2$ can be found in Ref.\cite{Lottini:2011zp}.

\section{Calculation of the observables}

The fermion action with $N_f=2$ and with a diquark source term $J$, 
necessary to study the diquark condensate, is
given by:
\beq
S_f=\bar{\psi}_1 M(\mu) \psi_1+\bar{\psi}_2 M(\mu) \psi_2
-J \bar{\psi}_1 (C \gamma_5) \tau_2 \bar{\psi}_2^{tr}
+J \psi_2^{tr} (C \gamma_5) \tau_2 \psi_1 \ ,
\eeq
where $M(\mu)$ is the standard Wilson fermion matrix at non-zero 
chemical potential and $C$ is the charge conjugation operator.
If we introduce the change of variables $\bar{\phi}= -\psi_2^{tr} C \tau_2 $,  
$\phi= C^{-1} \tau_2 \bar{\psi}_2^{tr}$, $\psi=\psi_1$, $\bar\psi=\bar\psi_1$, 
it is possible to rewrite the action as:
\beq
S_f= ( \bar\psi \bar\phi ) 
\left(
\begin{array}{cc}
M(\mu) & J \gamma_5 \\
-J \gamma_5 & M(-\mu) 
\end{array}
\right)
\left(
\begin{array}{c}
\psi \\
\phi
\end{array}
\right)
\equiv \bar\Psi \mathcal{M} \Psi \ .
\eeq

It is worth mentioning that 
$
\det{(\mathcal{M}^\dagger\mathcal{M})}=\left[ \det{(M^\dagger M+ J^2)} 
\right]^2
$, 
therefore we can take the square root analytically, {\it i.e.} there is no 
square root problem. The partition function becomes:
\beq
Z= \int dU d\bar{\Psi} d\Psi e^{-S_g-\bar\Psi \mathcal{M} \Psi} \ .
\label{pfnew}
\eeq
It is now easy to write down an expression for $n_q$ using 
the matrix $\mathcal{M}$:
\beq
n_q= \frac{T}{V_s} \frac{\partial \ln{Z}}{\partial \mu}
= \frac{T}{V_s} \sum_{\alpha,\beta} 
\langle - \bar\Psi_\alpha \left( \frac{\partial \mathcal{M}}
{\partial \mu} \right)_{\alpha,\beta} \Psi_\beta  \rangle
=
\frac{T}{V_s} \langle  \mbox{Tr} \left\{ 
\mathcal{M}^{-1} \frac{\partial \mathcal{M}} {\partial \mu} \right\}
\rangle \ .
\eeq
Moreover, from the definition of QNS we have:
\beq
\chi=\frac{\partial n_q}{\partial \mu}
=\frac{T}{V_s} \left\{ - \langle \left[ -\bar\Psi \frac{\partial 
\mathcal{M}}{\partial \mu} 
\Psi \right] \rangle ^2
+
\langle \left[ -\bar\Psi \frac{\partial \mathcal{M}}{\partial \mu} 
\Psi \right]^2 \rangle
+
\langle \left[ -\bar\Psi \frac{\partial^2 \mathcal{M}}{\partial \mu^2} 
\Psi \right] \rangle \right\} \label{chi3terms} \ .
\eeq
From this equation we can identify four different terms:
\bea
T1&=& - \langle \left[ -\bar\Psi \frac{\partial \mathcal{M}}{\partial \mu} 
\Psi \right] \rangle ^2 
= - \langle \mbox{Tr} \left[ \mathcal{M}^{-1}  \frac{\partial \mathcal{M}}
{\partial \mu} \right] \rangle ^2 \label{T1} \\
T2&=& + \langle \left[ -\bar\Psi \frac{\partial \mathcal{M}}{\partial \mu} \Psi 
\right]^2 \rangle_{disc}=\langle \mbox{Tr} \left[ \mathcal{M}^{-1}  
\frac{\partial \mathcal{M}}{\partial \mu}\right] \cdot \mbox{Tr} 
\left[ \mathcal{M}^{-1}  \frac{\partial \mathcal{M}}{\partial \mu}\right] 
\rangle \label{T2}  \\
C1&=& + \langle \left[ -\bar\Psi \frac{\partial \mathcal{M}}{\partial \mu} 
\Psi \right]^2 \rangle_{conn}=- \langle \mbox{Tr} \left[ \mathcal{M}^{-1}  
\frac{\partial \mathcal{M}}{\partial \mu} \mathcal{M}^{-1}  \frac{\partial 
\mathcal{M}}{\partial \mu}\right] \rangle \label{C1} \\
T3&=& + \langle \left[ -\bar\Psi \frac{\partial^2 \mathcal{M}}{\partial \mu^2} 
\Psi \right] \rangle= \langle \mbox{Tr} \left[\mathcal{M}^{-1} \frac{\partial^2
\mathcal{M}}{\partial \mu^2}  \right] \rangle \label{T3}  \ .
\eea

We see that from the second term of Eq.~(\ref{chi3terms}) we get two terms, 
namely $T2$ and $C1$, because there are two ways to contract the spinors.

The calculation of the traces is done by unbiased estimators, introducing 
$N_\eta$ complex noise vectors $\eta$ with the properties: 
$\langle \eta_x \rangle =0$ and $\langle \eta_x \eta_y \rangle = \delta_{x y}$.
For example, the determination of the following trace, used for $T1$ and $T2$,
is based on the relation:
\beq
\mbox{Tr} \left[ \mathcal{M}^{-1}  \frac{\partial \mathcal{M}}{\partial \mu}
\right] =
\frac{1}{N_\eta}\sum 
\eta^*_{x \alpha i}\left( \frac{\partial \mathcal{M}}{\partial \mu} 
\right)_{x \alpha i ; y \beta j} \mathcal{M}^{-1}_{y \beta j ; z \gamma k}
\eta_{z \gamma k} \ .
\label{trmmdm}
\eeq
Because for $T2$ we need two independent source vectors we refer to this term
as disconnected term; the other three terms need only one source vector and
we call them connected terms.

\section{Some numerical issues}

The source vector used to determine the traces can characterised by 
different noise distributions.
The standard method is based on the introduction of gaussian complex noise 
vector but it is possible also to use a $Z_2$ complex noise vector 
(see Ref.~\cite{Dong:1993pk} where the potential advantage are discussed). 
In the $Z_2$ case, the complex noise vectors $\eta$ 
takes one of the four values $\left\{ \pm 1, \pm i \right\}$, chosen 
independently with equal probability.

In Table~\ref{tablenoise} we present an example where we used 
only three noise vectors, with the following parameters: 
$\beta=1.70$, $\kappa=0.178$, $j=0.04$, $\mu=0.25$,  $8^3 \times 16$.
\begin{table}[htbp]
\begin{center}
\begin{tabular}{|l|l|l|l|l|l|}
\hline
\phantom{.} & T1 & T2 & T3 & C1 & $\chi$  \\
\hline
Gaussian & 2.25(8.30)E-06 & 5.00(5.74)E-05 & 0.4010(27) & -0.3681(107) & 3.1(1.0)  \\
\hline
$Z_2$    & 6.3(10.0)E-06  & 7.4(49.9)E-06  &  0.3994(32) & -0.3655(66) & 3.17(52)  \\
\hline
\end{tabular}
\end{center}
\caption{Gaussian vs $Z_2$ noise vectors.}
\label{tablenoise}
\end{table}
From this simple analysis, we can see that there is no any particular 
advantage in using a $Z_2$ noise, so in the following we continue to use only
gaussian noise distributions.

We also tried to see what happen when we increase the number of noise vectors;
in Table~\ref{tablenumbersources} we present an example with the 
same parameters used above.
\begin{table}[htbp]
\begin{center}
\begin{tabular}{|l|l|l|l|l|l|}
\hline
 noise vectors & T1 & T2 & T3 & C1 & $\chi$  \\
\hline
3      & 2.25(8.30)E-06 & 5.00(5.74)E-05 & 0.4010(27) & -0.3681(107) & 3.1(1.0)  \\
\hline
300    & 1.50(17)E-05  & 1.53(12)E-05 & 0.4013(14)  & -0.3612(20) & 4.04(11)E-02 \\
\hline
\end{tabular}
\end{center}
\caption{The effect of different number of noise vectors.}
\label{tablenumbersources}
\end{table}
From this result it is evident how increasing the number of noise vectors 
has a strong effect mainly on observables with a small value; the effect 
is very limited if the observable is clearly different from zero.
\begin{figure}[ht]
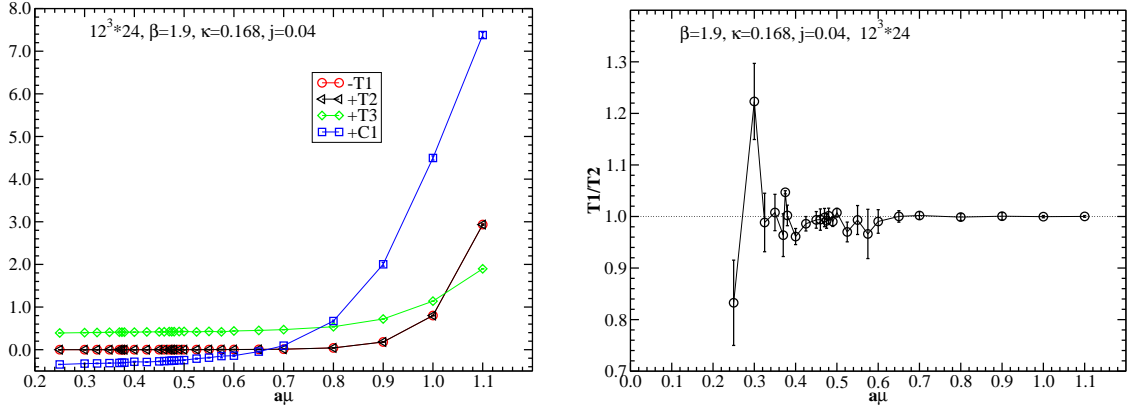

\center
\vspace{5mm}
\includegraphics[width=7cm]{plots/plot-terms.eps}
\hspace{5mm}
\includegraphics[width=7.2cm]{plots/plot-t1vst2.eps}
\caption{(Left) The four terms of the QNS are plotted against the
chemical potential. (Right) The ratio between T1 and T2 is plotted against 
the chemical potential.}
\label{plot1}
\end{figure}
In Fig.~\ref{plot1} (Left) we plot the four terms (note the sign of $T1$)
vs the chemical potential. The connected contribution $C1$ gives
clearly an important contribution either at low and high values of the 
chemical potential; note the changing of sign around $a\mu \approx 0.66$.
The terms $T1$ and $T2$ are equal in magnitude but with opposite sign,
therefore their contribution cancels everywhere, Fig.~\ref{plot1} (Right); 
in other words the variance of the quantity in
Eq.~(\ref{trmmdm}) is equal to zero. We see very strong fluctuations,
apparently not compatible with one, for $\mu < \mu_0$, then smoothed 
fluctuations comparable with one until $\mu_D$ and after this a stable
ratio equal to one.

\section{Numerical results}

In this contribution we are going to show some results obtained for $\beta=1.9$
and $\kappa=0.168$. From previous works, see Ref.~\cite{Hands:2010gd} and 
the references therein, we know that for these parameters $m_\pi=0.68(1)$,
therefore $a\mu_0 \approx 0.34$; moreover, there are signals of a BEC phase
for $a\mu \lesssim 0.45$.

We obtained interesting results comparing the QNS with the other 
observables; here we want to stress an interesting 
effect we have observed studying the system at different temperatures.
Note that the results we are presenting should eventually be 
extrapolated to $j=0.00$.

\begin{figure}[ht]
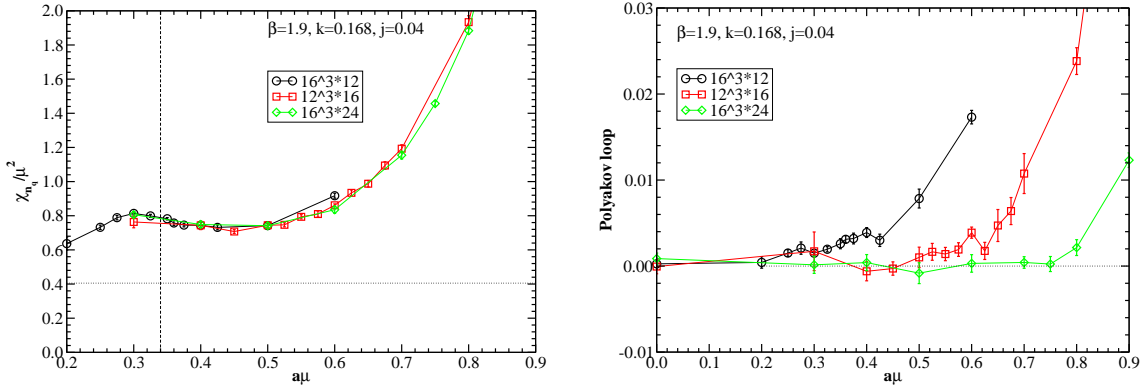

\center
\vspace{5mm}
\includegraphics[width=7.2cm]{plots/plot-suscett_vsmusq.eps}
\hspace{5mm}
\includegraphics[width=7.2cm]{plots/plot-polyak2.eps}
\caption{(Left) Ratio $\chi/\mu^2$ versus $\mu$.
The horizontal dotted line marks the SB value $4/\pi^2$ and the vertical 
dashed one marks the position of $\mu_0$. (Right) Polyakov 
loop versus $\mu$.}
\label{plot2}
\end{figure}

In Fig.~\ref{plot2} (Left) we plot the ratio $\chi/\mu^2$, for three different 
temperatures, versus the chemical potential.  For an ideal gas of
quarks this ratio would be a constant, see
Eq.~(\ref{nchinSB}), and we see that
a plateau is actually present for $a\mu \lesssim 0.55$; after this value
we can see a sharp increase of the QNS.
Moreover, it is evident from this plot that the QNS is quite independent from 
the temperature, {\it i.e.} we do not see any drastic deviations in the 
behaviour of the three curves increasing $\mu$.
This is in contrast with the Polyakov loop behaviour that shows deconfinement 
for three different values of the chemical potential, correspondingly at 
the three different temperatures, Fig.~\ref{plot2} (Right).

It is instructive to compare our lattice numerical results
with the equations corresponding to Eq.~(\ref{nchinSB}) but taking in account 
the finite volume and the lattice discretisation.
In Ref.~\cite{Hands:2006ve}, see Eq.~(26), the expression for
the quark number density $n_q^{SBL}$ for free Wilson fermions
on the lattice is presented.  It is then easy to obtain $\chi^{SBL}$.

\begin{figure}[ht]
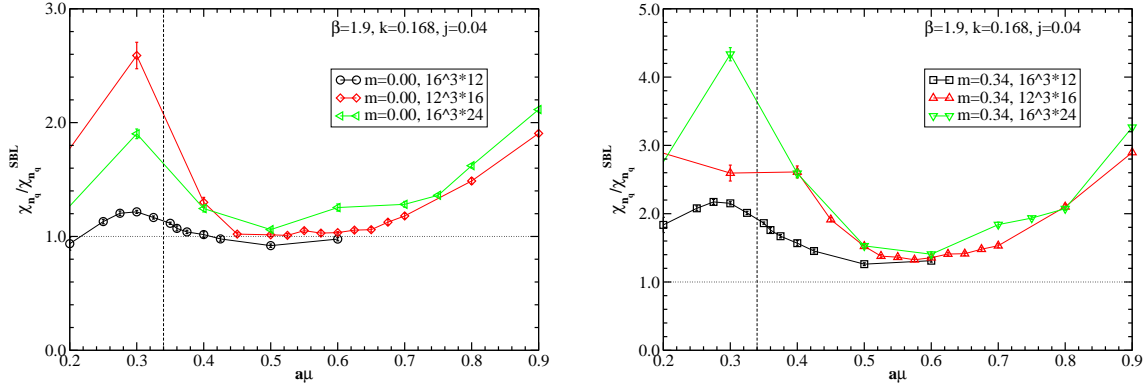

\center
\vspace{5mm}
\includegraphics[width=7.2cm]{plots/plot-suscett_2.eps}
\hspace{5mm}
\includegraphics[width=7.2cm]{plots/plot-suscett_3.eps}
\caption{The ratio between the measured QNS and the ideal value 
for \emph{lattice} free fermions for two values of the fermion mass: 
$m=0.00$ and $m=0.34$.
The vertical dashed lines mark the position of $\mu_0$.}
\label{plot3}
\end{figure}

In Fig.~\ref{plot3} we plot the ratio between the measured QNS and $\chi^{SBL}$ 
for two values of the fermion mass: in the determination of $\chi^{SBL}$
we have to fix a value for the mass of the free fermion; unfortunately we do 
not know this value, therefore we consider massless fermions (note that this 
is the same limit used in Eq.~(\ref{nchinSB})) and a value of the order
of $m_\pi/2$.
In this case we observe a different behaviour for $a\mu \lesssim 0.45$, 
reminiscent of the BEC phase, followed by a flat region compatible with 
ratio one,
{\it i.e.} the system is behaving as free fermions, and then again we see
an increase for higher values of $\mu$. 
These plots again confirm the above scenario: we do not see any abrupt 
change for QNS, for any of the values of $\mu$,  where instead the Polyakov 
loop becomes different from zero.

Note that the QNS is often taken as an alternative 
signal for deconfinement in lattice studies of the thermal QCD transition 
{\it i.e.} there is a strong connection between $\chi$ and Polyakov 
Loop, see Refs.~\cite{Aoki:2006br, Bazavov:2009zn, Hands:2010vw}.
Our results shows that this relation is not replicated at low temperature 
and high density.

\section{Conclusion}

In this work we have studied the quark number susceptibility
at non-zero density and low temperature in the case of two color QCD with
two flavours. We have studied the contribution of the different terms,
connected and disconnected, which contribute to it and finally we have
shown its behavior at finite temperature. Surprisingly, we have seen that
in this case there is no relation between the quark number susceptibility
and the Polyakov loop, {\it i.e.} in this context it cannot be used as
a signal for deconfinement as it is usually done at finite temperature and
zero quark number density.
This observation could suggest that the deconfinement transition 
is not characterised by a liberation of additional degrees of freedom;
if this phenomenon is exclusive to two color QCD or related
to the non small ratio $m_\pi/m_\rho = 0.80(1)$ or, somehow, connected 
to the presence of a quarkyonic phase, is under study.
Clearly, there is much to learn about deconfinement from studying 
a new physical environment.


\end{document}